\documentclass[12pt]{article}
\usepackage{pst-plot,epsf}
\newcommand{\al }{\alpha}
\newcommand{\gamu }{\gamma_\mu}
\newcommand{\epm}{e^+e^-}

\newcommand{\be}{\begin{equation}}
\newcommand{\ee}{\end{equation}}
\newcommand{\ba}{\begin{eqnarray}}
\newcommand{\ea}{\end{eqnarray}}
\newcommand{\bea}{\begin{eqnarray*}}
\newcommand{\bet}{\begin{center} \begin{tabular}}
\newcommand{\eea}{\end{eqnarray*}}
\newcommand{\ent}{\end{tabular} \end{center}}
\newcommand{\bb}{\begin{thebibliography}}
\newcommand{\eb}{\end{thebibliography}}
\newcommand{\ci}[1]{\cite{#1}}

\newcommand{\lab}[1]{\label{#1}}
\newcommand{\ra}{\rightarrow}

\newcommand{\bit}{\begin{itemize}}
\newcommand{\eit}{\end{itemize}}

\newcommand{\veps}{\varepsilon}

\newcommand{\crn}{\nonumber \\}
\newcommand{\nn}{\nonumber}

\newcommand{\gv}{\mbox{GeV}}

\newcommand{\MOM}{${\mathrm{MOM}}$ }

\newcommand{\MSb}{$\overline{\mathrm{MS}}$ }
\newcommand{\MSbm}{\overline{\mathrm{MS}} }

\begin{document}

\parindent 0mm
\parskip 2mm
\renewcommand{\arraystretch}{1.4}
\thispagestyle{empty}
\hfill {\sc DESY 98-206} \qquad { } \par
\hfill December 1998 \qquad { }

\vspace*{15mm}

\begin{center}
\renewcommand{\thefootnote}{\fnsymbol{footnote}}
\begin{center}

{\Large {\bf Testing non--perturbative strong interaction effects\\ via
the Adler function}\vglue 10mm }

\vspace{1.5cm}

{\large
 S. Eidelman$^{a}$, F. Jegerlehner$^{b}$,
 A.L. Kataev$^{c}$\footnote{Partially supported by Landau--Heisenberg
program and Russian Foundation of Basic Research, \\ 
\hspace*{6mm} Grants N 96-01-01860, N 96-02-18897}, O. Veretin$^{b}$}

\vspace{1cm}
$^a$~~Budker Institute for Nuclear Physics~~~~~\\
630090 Novosibirsk, Russia

\vspace{5mm}

$^b$~~Deutsches Elektronen-Synchrotron DESY~~~~~ \\
Platanenallee 6, D--15738 Zeuthen, Germany\\

\vspace{5mm}

$^c$~~Institute for Nuclear Research of the Academy of Sciences of Russia~~~~~ \\
117312 Moscow, Russia\\
\end{center}

\textwidth 120mm
\begin{abstract}
Based on the compilation of the available $e^+e^-$ data~\ci{EJ}, we
present a non--perturbative estimation of the Adler function derived
from the electromagnetic current correlator, and compare it with
theoretical predictions from perturbative QCD (pQCD). The comparison
is presented for the Euclidean region where pQCD is supposed to work
best. We emphasize that such a comparison only makes sense if one
takes into account the exact mass dependence of the perturbative
predictions, which are available for the leading and next to leading
(two--loop) order. In order to have the correct physical mass
dependence in the evolution of the strong coupling as well, we utilize
the \MOM scheme $\beta$--function to two--loops calculated
recently~\ci{JT}. Three--loop effects, which are available as series
expansions for low\ci{ChKSsQ} and high\ci{ChHKShQ} momentum transfer,
are included by using Pad\'e improvement. We discuss possible
constraints on non--perturbative effects as suggested by the operator
product expansion.
\end{abstract}
\end{center}

\bigskip
\textwidth 170 mm
\newpage

\renewcommand{\thefootnote}{\arabic{footnote}}

{\bf 1. Introduction}

There are different ways of comparing QCD theoretical predictions with
available time-like experimental data for the $e^+e^-$ total
cross-section.  Among the proposed methods are the global duality sum
rules with different smearing functions
\cite{PQW,Barnet,SVZ,DeRafael} as well as the local duality sum
rules \cite{CKT,Shankar}. The Borel sum rules have already been used
before to analyze the low-energy $e^+e^-$ data \cite{E} and gave the
reasonable value of the gluon condensate $<G^2>$ and a rather low
value of the parameter $\Lambda_{\overline{MS}}^{(3)}$, which turns
out to be in contradiction with the current world average value of
$\alpha_s(M_Z)=0.119\pm 0.002$~\cite{PDG}.

In view of the improved understanding of the influence of the
perturbative QCD corrections (both massless \cite{ChKT,GKL} and
massive ones \cite{ChKSsQ,ChHKShQ,GKLM}) and the appearance of the
compilation of the available $e^+e^-$ data \cite{EJ} it is worthwhile
to study the problem of the comparison of the experimental data with
the theoretical predictions at the new level.
  
In this our work we will follow the idea originally proposed in
Ref.\cite{Adler} and implemented in part in Ref.\cite{RG} and will use
for the comparison with theory the experimental data for the Adler
function, defined through the following dispersive relation
\begin{equation}
D(Q^2)=Q^2\int_{4 m_{\pi}^2}^{\infty}\frac{  R^{\rm exp}(s)}{(s+Q^2)^2}ds\;\;,
\label{Adler}
\end{equation}
where $Q^2=-q^2$ is the squared Euclidean momentum transfer, $s$ is the
center of mass energy squared for hadron production in
$\epm$--annihilation,
\begin{equation}
R(s)=\frac{\sigma_{\rm tot}(e^+e^-\rightarrow \gamma^*\rightarrow
  {\rm hadrons})}{\sigma(e^+e^-\rightarrow \gamma^*\rightarrow \mu^+\mu^-)}
\end{equation}
the hadronic to leptonic cross section ratio with
$\sigma(e^+e^-\rightarrow
\gamma^*\rightarrow \mu^+\mu^-)=\frac{4\pi\alpha^2}{3s}$, 
$\alpha=e^2/4\pi$ being the QED fine structure constant.

The attractive features of this procedure were advocated from different
points of view in Ref.\cite{AK} and Ref.\cite{Raf}. 

{\bf 2. Calculations of the Adler function}

In this note we reconsider the question of the reliability of
pQCD predictions at lower energies (down to about 1
GeV) by calculating the Adler function associated with the hadronic part of
the electromagnetic current
\bea 
J^{\gamma}_\mu \,=\,
\sum_{q}\;Q_q\, J_{\mu\:V}^q \,=\, \sum_{i}\;(
\frac{2}{3} \bar{u}_i \gamu u_i -\frac{1}{3} \bar{d}_i \gamu d_i)\;.\crn
\eea 
By $u_i$ and $d_i$ we denote the upper and lower components of
the quark weak--isodoublets, respectively ($i$ is the family index).   
\bea 
J^{q}_{\mu\:V} &=& \bar{q} \gamu q
\eea
is the conserved flavor diagonal vector current of the quark $q$ with
charge $Q_q$ and mass $m_q$.

We first define the photon vacuum polarization amplitude 
$\Pi'_{\gamma} (q^2)$ by
\ba 
\Pi^{\gamma}_{\mu \nu}(q) &=& i
\int d^4 x e^{iqx} <0|T J^\gamma_\mu\; (x)\; J^\gamma_\nu\;(0)\;|0>
\crn &=& -\left(q^2\, g_{\mu \nu}- q_\mu q_\nu\right)\;
\Pi'_{\gamma}\;(q^2)\;\;.
\label{CC}
\ea 
By $\Pi'_{V}$ we denote the corresponding amplitude for the vector
current $J^{q}_{\mu\:V}$. 

The Adler function is defined as the derivative 
\ba
D(-s)=-(12\pi^2)\,s\,\frac{d\Pi'_{\gamma}\,(s)}{ds}\;.
\label{Adef}
\ea 
We may write 
\ba 
D(-s)=\frac{3\pi}{\alpha} s\frac{d}{ds}\Delta \alpha_{\mathrm{had}}(s)
\label{DD}
\ea 
where  
\be 
\Delta \al _{\mathrm{had}}(q^2) = -\frac{\al
q^2}{3\pi}\,{\mathrm{Re}} \int_{4m_{\pi}^2}^{\infty}
ds\,\frac{R(s)}{s(s-q^2-i\veps)}  
\label{DI}
\ee 
is the hadronic contribution to the shift of the fine structure
constant at scale $q^2$. A careful estimate of this function using the
existing experimental data was given some time ago in Ref.~\cite{EJ}.
Experimental data for $R(s)$ may be used up to $E_{\mathrm{cut}}=40$
GeV, for higher energies $\gamma Z$ mixing would complicate the
analysis. The high energy tail is calculated by using pQCD. Up to
three--loops results are available for massive quarks~\cite{ChK}. The
four--loop corrections are known in the approximation of massless
quarks~\cite{GKL}.

In the present note we are going to evaluate the Adler function by the
same procedure. It is important that we work in the Euclidean region
$Q^2=-q^2>0$ where pQCD is supposed to work best.

Before presenting the non-perturbative result, which follows from the
estimate of the dispersion integral, we briefly discuss the analytic
results which were obtained from perturbative QCD and the operator product
expansion (OPE). We write
\ba
D(Q^2)=D^{(0)}(Q^2)+D^{(1)}(Q^2)+D^{(2)}(Q^2)+\cdots+D^{\mathrm{NP}}(Q^2)
\ea
where $D^{(n)}(Q^2)$ is the $n$--th order pQCD contribution
proportional to $(\alpha_s/\pi)^n$ and $D^{\mathrm{NP}}(Q^2)$ is the
non-perturbative part which will be specified below. 

We first consider the pQCD contributions.
For the representation of the explicit expressions we use the variables
\ba                               
y=4m_f^2/s \;\;,\;\;\;  \xi=\frac{\sqrt{1-y}-1}{\sqrt{1-y}+1} 
\label{xivar}
\ea
where $\xi$ is taking values $0 \leq \xi \leq 1$  for $ s \leq 0\;.$ 

The \MSb subtracted transverse photon vacuum polarization function 
$\Pi'_\gamma(q^2)$, which defines the full photon propagator as
\bea
-g_{\mu\nu}\frac{i}{q^2} \frac{1}{1+e^2  \Pi'_\gamma(q^2)}
\eea
at the one--loop level is given by
\bea
e^2 \Pi'_\gamma(q^2)=\frac{\alpha}{3\pi} \sum_f Q_f^2 N_{cf}
\left(\ln \frac{\mu^2}{m_f^2}+ G  \right)
\eea
with $\mu$ the \MSb scale parameter,
$N_{cf}$ the color factor and
\bea
G=\frac{5}{3}+y+\left(1+\frac{y}{2} \right)\:\sqrt{1-y} 
\ln \xi  \;.
\eea
The corresponding one--loop contribution to $\Delta \alpha$ is 
\bea
\Delta \alpha(q^2)=e^2\left(\Pi'_\gamma(0)-\Pi'_\gamma(q^2)\right)
\eea
while for the one-loop Adler function we obtain
\ba
D^{(0)}(Q^2) =
\sum_f Q_f^2 N_{cf} H^{(0)}
\label{D1}
\ea
with
\ba
H\equiv (12\pi^2)\,\dot{\Pi}'_V\;,\;\;\;\dot{\Pi}'_V\equiv -s\,d\Pi'_V/ds\;,
\ea
in terms of the vector current amplitude $\Pi'_V$. Explicitly we have
\bea
H^{(0)}= y\frac{dG}{dy}= 1+ \frac{3y}{2}- 
\frac{3y^2}{4}\frac{1}{\sqrt{1-y}}\ln \xi \;. 
\eea
Asymptotically, we find
\bea
H^{(0)}\ra \left \{ \begin{array}{lr}
\frac{1}{5}\frac{Q^2}{m_f^2}-
\frac{3}{70}\left(\frac{Q^2}{m_f^2}\right)^2+
\frac{1}{105}\left(\frac{Q^2}{m_f^2}\right)^3+
\cdots &
Q^2\ll m_f^2 \\
1-6\frac{m_f^2}{Q^2}-
12\left(\frac{m_f^2}{Q^2}\right)^2 \ln\frac{m_f^2}{Q^2}+
24\left(\frac{m_f^2}{Q^2}\right)^3 \left(\ln\frac{m_f^2}{Q^2}+1\right)+\cdots & 
Q^2\gg m_f^2  
		    \end{array} \right.
\eea
and this behavior determines the quark parton model (QPM) (leading
order QCD) property of the
Adler function: heavy quarks ($m_f^2\gg Q^2$) decouple like
$Q^2/m_f^2$ 
while light modes ($m_f^2\ll Q^2$) contribute $Q_f^2 N_{cf}$ to $D^{(0)}$.
Using the two--loop QCD vacuum polarization functions given in the 
appendix we obtain for the two--loop contribution to
the Adler function
\ba
D^{(1)}(Q^2)=\frac{\al_s(Q^2)}{\pi} \sum_f Q_f^2 N_{cf} 
\,H^{(1)}
\label{Atwo}
\ea
where $H^{(1)}=(12\pi^2)\,
\dot{\Pi}^{'(2)}_V(-Q^2,m^2_f)$.

The full massive three--loop vector current correlator $\Pi^{'(3)}$
is not yet known, however, a sufficient number of terms were
calculated both for the small mass expansion~\cite{ChHKShQ} as well as
for the heavy mass expansion~\cite{ChKSsQ}. By taking the derivative
(\ref{Adef}) we obtain the contribution to the Adler function
\ba
D^{(2)}(Q^2)=\left(\frac{\al_s(Q^2)}{\pi}\right)^2 \sum_f Q_f^2 N_{cf} 
\,H^{(2)}
\label{Athree}
\ea
where
$H^{(2)}=(12\pi^2)\,
\dot{\Pi}^{'(3)}_V(-Q^2,m^2_f)$.

The convergence of the series expansions break down at Euclidean
$Q^2=4m^2$, right in the region where mass effects are most important,
and we use a conformal mapping together with Pad\'e
improvement~\cite{FT} in order to obtain results useful for the
present analysis. For details we refer to a brief discussion in the
appendix.

A way to parametrize non-perturbative (NP) effects away from resonances,
as applicable here, is prescribed by the OPE of the current
correlator~(\ref{CC}), valid for large enough $Q^2$. This part is due
to non--vanishing gluon and light quark condensates~\cite{SVZ}. They
yield the leading power corrections
\ba
\label{NP}
D^{\mathrm{NP}}(Q^2) &=& \sum\limits_{q=u,d,s} Q_q^2 N_{cq}\,(8\pi^2) \crn 
&\cdot& \left[\frac{1}{12} 
\left(1-\frac{11}{18}\frac{\alpha_s(\mu^2)}{\pi}\right) 
\frac{<\frac{\alpha_s}{\pi} G G>}{Q^4} \right. \crn
&+&2\,\left(1+\frac{\alpha_s(\mu^2)}{3\pi} +\left(\frac{47}{8}-\frac34
l_{q\mu}
\right)\left(\frac{\alpha_s(\mu^2)}{\pi}\right)^2\right)
\frac{<m_q \bar{q}q>}{Q^4} \\
&+&\left. \left(\frac{4}{27}\frac{\alpha_s(\mu^2)}{\pi}
+\left(\frac{4}{3}\zeta_3-\frac{88}{243}-\frac13 l_{q\mu}
\right)
\left(\frac{\alpha_s(\mu^2)}{\pi}\right)^2\right)
\sum\limits_{q'=u,d,s} \frac{<m_{q'} \bar{{q'}}{q'}>}{Q^4}\,+\, \cdots
\right] \nn 
\ea 
where $l_{q\mu}\equiv\ln(Q^2/\mu^2)$. $<\frac{\alpha_s}{\pi} G G>$ and
$<m_q \bar{q}q>$ are the scale-invariantly defined condensates. The
terms beyond leading order in $\alpha_s$ have been calculated from the
results which were obtained in
Refs.~\cite{NPho},~\cite{ChGS},~\cite{SCh} and~\cite{ST90}. The terms
beyond the leading order in $\alpha_s$ are unimportant because
experimental data only yield rough constraints on the condensates and
we include them for completeness only. Sum rule estimates of the
condensates yield~\cite{SVZ,Nar89}
\bea
<\frac{\alpha_s}{\pi} G G>\simeq (0.336 \cdots 0.442\; \gv
)^4\;\;{\mathrm{and}}\;\; <m_q \bar{q}q >\simeq \left\{ 
\begin{array}{l}
-(0.086 \cdots 0.111\; \gv )^4\;\;{\mathrm{for}}\;\;q=u,d \\
-(0.192 \cdots 0.245\; \gv )^4\;\;{\mathrm{for}}\;\;q=s\;.
\end{array} \right. 
\eea
with rather large uncertainties. The values assumed for the quark
condensates are based on the recent estimate\cite{DoNa} $< \bar{q}q
>\simeq -\left(235 \pm 27 \;\;{\mathrm MeV}\right)^3$ and the quark
masses $m_u\simeq m_u \simeq (m_u+m_d)/2 \simeq 7.2 \pm 1.1$ MeV and $m_s
\simeq 175\pm 25$ MeV~\cite{Gasser}, all at scale 1 GeV. 
Eq.~(\ref{NP}) will be evaluated numerically at $\mu^2=Q^2$.

{\bf 3. Numerical results and conclusions}

For numerical studies we used the following pole quark masses~\cite{PDG}
$$m_u \sim m_d \sim m_s \sim 0\;;\;\;m_c=1.55 \gv \;;\;\;m_b=4.70 \gv
\;;\;\;m_t=173.80 \gv \;.$$ For the strong interaction coupling we
take $\alpha_{s\; {\small \MSbm}}^{(5)} = 0.120 \pm 0.003$ at the scale
$M_Z$=91.19 GeV~\cite{LEP}.  Up to two--loop we use the exact mass
dependence (see e.g.~\cite{FR}) of the amplitudes as well as of the
exact mass dependent two--loop $\overline{\alpha}_s(Q^2)$ in the gauge
invariant background--field \MOM renormalization scheme, presented
recently in Ref.~\cite{JT}. In the transition from the \MSb to the
\MOM scheme we adapt the rescaling procedure described in~\cite{JT},
such that for large $\mu$
\bea
\overline{\alpha}_s((x_0\mu)^2)=\alpha_s(\mu^2)+0+ O(\alpha_s^3)\;.
\eea
This means that $x_0$ is chosen such that the couplings coincide to
leading and next--to--leading order at asymptotically large
scales. Numerically we find $x_0\simeq 2.0144$. Due to this
normalization by rescaling the coefficients of the Adler--function
remain the same in both schemes up to three--loops.

The numerical evaluation of the dispersion integral (\ref{Adler}) is
based on the compilation of data

\rput{90}(8.7,-3.0){\scalebox{0.65 0.65}{%
\epsfbox{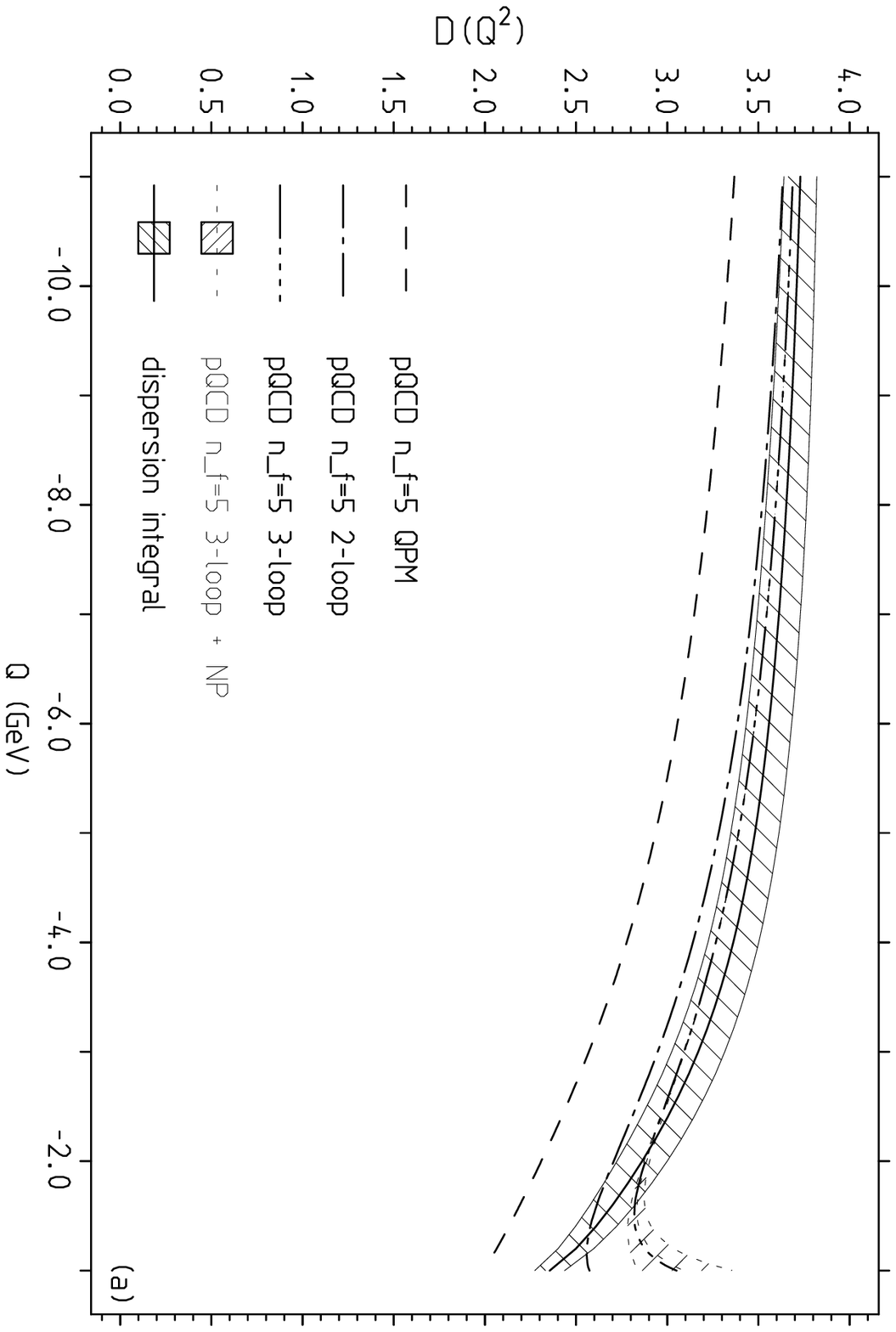}}}

\vspace*{7cm}

\rput{90}(8.7,-4.0){\scalebox{0.65 0.65}{%
\epsfbox{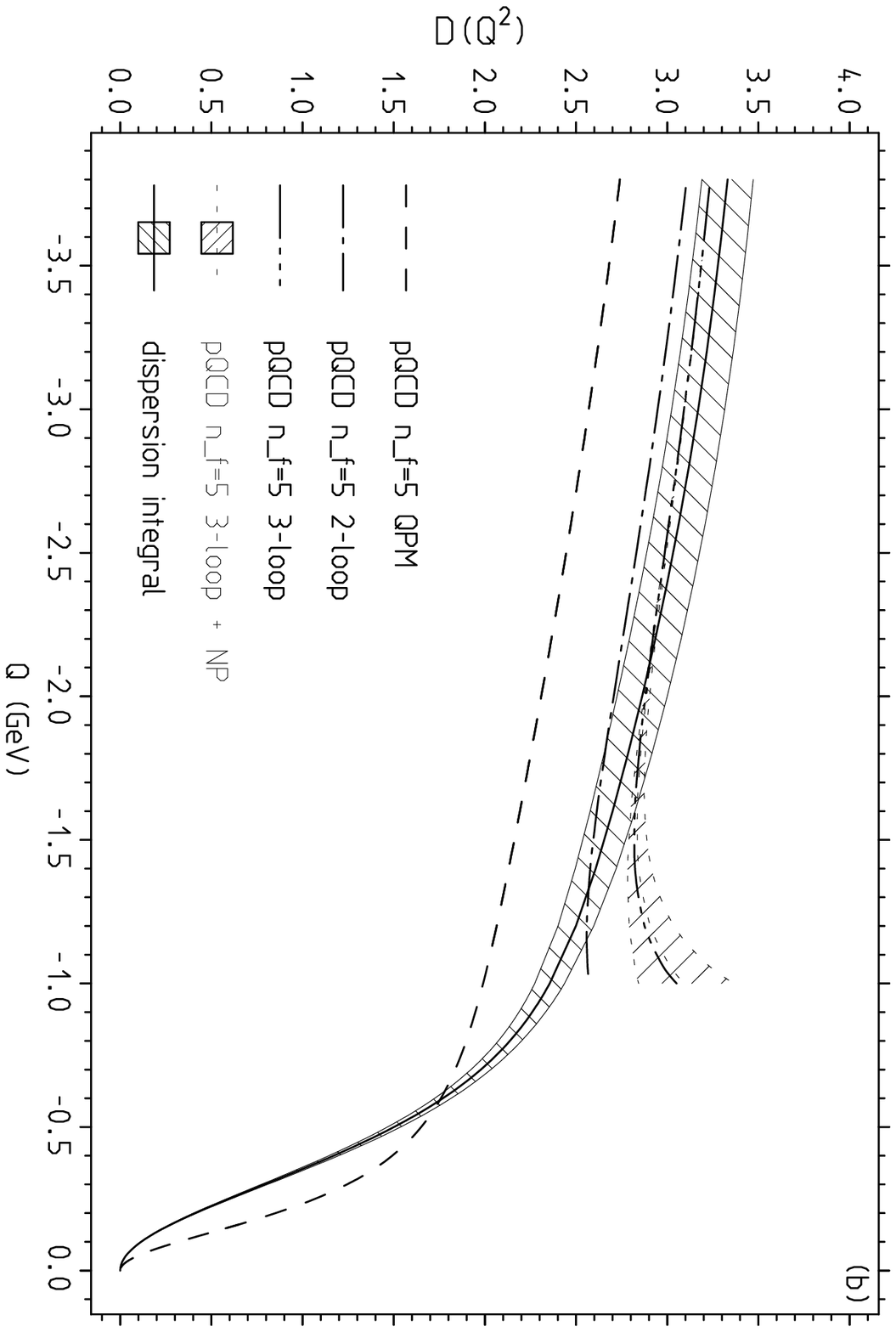}}}

\vspace*{8.9cm}

\begin{center}
\begin{minipage}[h]{15.2cm} \baselineskip 12truept \noindent
\small{{Figures~1a and~1b: We show the ``experimental'' Adler function 
together with pQCD predictions and power corrections (NP) in the
region below about 10 GeV (a) and separately for the low energy tail
(b). The negative sign chosen for $Q$ should remind the reader that
we are considering space--like momentum transfer. The shaded area
represents the $\pm 1 \sigma$ band obtained from the data. A similar
error band is shown for the uncertainties in the power corrections.
}}
\end{minipage}
\end{center}

discussed in~\cite{EJ}. Here we will
not use all the data up to 40 GeV, however, still take a conservative
attitude, and replace data by pQCD results only where it is obviously
safe to do so.  Accordingly, in the regions from 4.5 GeV to
$M_{\Upsilon}$ and above 12 GeV we use pQCD, including the massive
results available up to three--loops~\cite{ChHKShQ} and the four--loop
contribution in the massless approximation~\cite{GKL}.  The
"experimental" behavior of the $D$-function is displayed in Figs.~1(a)
and~1(b), together with the theoretical predictions. 
The data and the error analysis utilized in the evaluation of 
(\ref{Adler}) are described in Ref.~\cite{EJ}. Both statistical 
and systematic errors of the data were taken into account. Note that 
the meaning of the 1$\sigma$ error band differs from the usual ``one 
standard deviation'' since its points at different $Q$ are highly 
correlated. The origin of the uncertainty is shown in Table~1.\\[-7mm]
\begin{center}
Table~1: Origin of uncertainties for 
the ``experimental'' Adler function at $Q$=2.5 GeV and $M_Z$.
\begin{tabular}{|c||rrr||rrr|}
\hline
     & $D(2.5\,\gv)$ & rel. err. & abs. err.
& $D(M_Z)$ & rel. err. & abs. err. \\
\hline
          Resonances:           &   .688 (.025) &   3.6 \% &   0.8 \% 
&   .004 (.000) &   5.2 \% &   0.0 \% \\
         $E<M_{J/\psi}$          &  1.068 (.127) &  11.9 \% &   4.2 \% 
&   .002 (.000) &  14.9 \% &   0.0 \% \\
    $M_{J/\psi}<E<3.6$ GeV      &   .178 (.035) &  19.9 \% &   1.2 \% 
&   .001 (.000) &  19.8 \% &   0.0 \% \\
  $3.6$ GeV $<E<M_{\Upsilon}$   &   .850 (.055) &   6.4 \% &   1.8 \% 
&   .032 (.002) &   7.0 \% &   0.1 \% \\
  $M_{\Upsilon}<E<$ 12 GeV      &   .088 (.008) &   8.7 \% &   0.3 \%
&   .024 (.002) &   9.0 \% &   0.1 \% \\
       $E < 12$ GeV data        &  2.871 (.146) &   5.1 \% &   4.8 \%
&   .063 (.003) &   4.9 \% &   0.1 \% \\
    12 GeV $  < E$ QCD          &   .162 (.001) &   0.3 \% &   0.0 \% 
&  3.755 (.006) &   0.2 \% &   0.2 \% \\
           total                &  3.033 (.146) &   4.8 \% &   4.8 \% 
&  3.818 (.007) &   0.2 \% &   0.2 \% \\
\hline
\end{tabular}
\end{center}

At $M_Z$ $D(Q^2)$ is completely dominated be the perturbative high
energy tail, the uncertainty is the one of pQCD. As $Q^2$ decreases 
the dominating region moves to lower energies. At 2.5 GeV contributions
to $D(Q^2)$ are spread over all energies below $M_\Upsilon$ and
uncertainties from the region below $M_{J/\psi}$ are dominating. Errors
at 2.5 GeV are about a factor 20 larger and dominated by systematic errors of
the $\epm$ data. 

The agreement between pQCD and the ``data'' looks rather good, but only
after inclusion of the 3--loop contribution. If we had not include the
3--loop term it would appear that there is some room for the
non--perturbative power corrections (\ref{NP}). After adding the
3--loop result there is no region left where the power corrections,
with the above given values for the parameters lead to an
improvement. So, the comparison of the perturbative predictions with
the "data" looks quite impressive after inclusion of the 3-loop
massive contribution. The inclusion of the top ($n_f$=6) does not
alter the results at the energies shown. We can not see a region of
intermediate $Q^2$ where adding the non-perturbative power corrections
(NP) does improve the agreement of 3-loop pQCD predictions with the
presented "data". It also does not look possible to estimate in a
convincing way the leading coefficient $<\frac{\alpha_s}{\pi} G G>$.
Due to the power behavior, these kind of corrections drop very fast at
larger energies, above about 2 GeV, while they grow very fast at lower
energies, below about 0.75 GeV, which signals the breakdown of the
applicability of the expansion. In fact, including the $1/Q^{6}$ terms
would lead to a large contribution of opposite sign at low $Q^2$. The
main problem is that these corrections become substantial only at
energies where the perturbation expansion can no longer be trusted.

It should be noted that in the Euclidean region there are no
``threshold steps'' as they are suggested if the \MSb scheme is
applied. In the \MSb scheme effective theories with different number
of light flavors must be matched at the different scales set by the
quark masses~\cite{Decoupling} in order to restore decoupling of heavy
flavors. The physical behavior is illustrated in Fig.~2. 

\rput{90}(8.7,-3.0){\scalebox{0.65 0.65}{%
\epsfbox{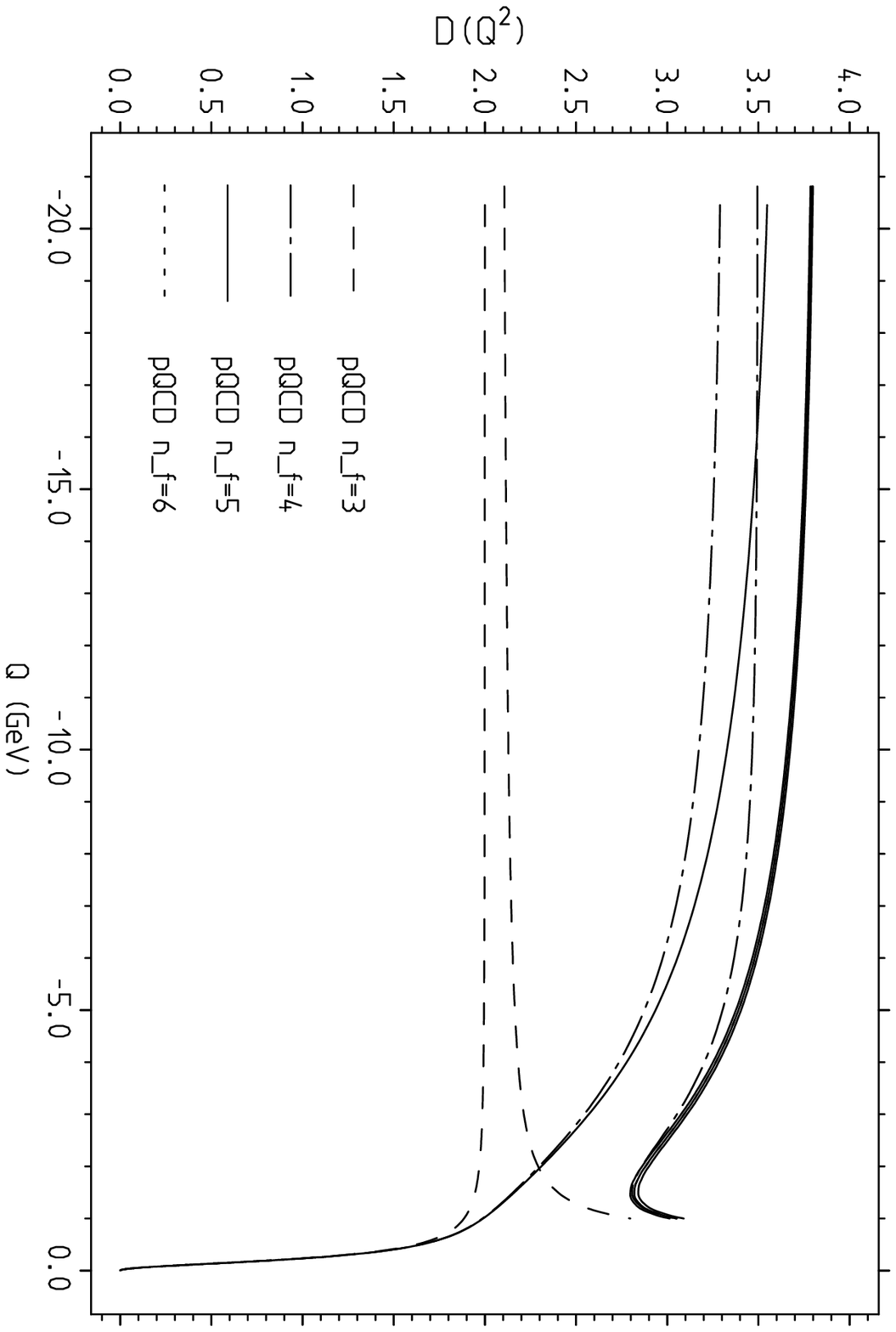}}}

\vspace*{7.9cm}

\begin{center}
\begin{minipage}[h]{15.2cm} \baselineskip 12truept \noindent
\small{{Figure~2: The flavor dependence of the Adler function in pQCD. 
For each $n_f=3 \cdots 6$ two curves are shown: one for the QPM (lower
of the two) and the other for pQCD at three--loops (upper of the
two). The cases $n_f$ =5 and $n_f$ = 6 are graphically indistinguishable at
the energies shown. The thickness of the $n_f$ = 5 $O(\alpha_s^2)$
``line'' illustrates the small theoretical uncertainty obtained from
the variation $\alpha_{s\; {\small \MSbm}}^{(5)} = 0.120 \pm 0.003$ at
the scale $M_Z$=91.19 GeV.  }}
\end{minipage}
\end{center}

In particular at low momenta the \MSb prescription does not provide a
good description of the physical behavior and hence its applicability
in the Euclidean region is worth a critical inspection. In this spirit
we also would expect an improvement of the QCD description of deep
inelastic scattering in cases where different effective numbers of
flavors $n_f$ have to be considered~\cite{BvN}.

Note that the 3--loop term is quite large and higher orders still
could give sizable contributions. The massless 4--loop contribution
may be estimated as follows: in massless pQCD we have the familiar 
result~\cite{GKL}
$R(s)=3\sum_f Q_f^2 
\times \left(1+a+c_1 a^2+c_2 a^3 +\cdots \right)$
with $a=\alpha_s (s)/\pi$, $c_1=1.9857-0.1153 n_f$ and
$c_2=-6.6368-1.2002 n_f - 0.0052 n_f^2 -1.2395\: (\sum Q_f)^2/(3 \sum
Q_f^2)$. In this approximation $R(s)$ is a constant if we discard the
running of $\alpha_s $ and we can evaluate (\ref{Adler}) obtaining
$D(Q^2)\sim R(Q^2)$ at large $Q^2$. Numerically the 4--loop term
proportional to $c_2$ amounts to $-0.06$\% at 100 GeV and increases to
about $-0.5$\% at 2.5 GeV. At the low energy this estimate is of
course not reliable because we do not expect the approximation to
work. Nevertheless in may indicate the possible size of the missing
higher order corrections.  Due to the increase of the running strong
coupling the higher order predictions become unreliable at lower
energies. Once we observe that the {\sf functional dependence}
deviates from that of the data, which happens below about 2.5 GeV,
pQCD is not reliable any longer. The large uncertainties in the data
(see Table~1) on the one hand and the beginning failure of pQCD on the
other hand, unfortunately do not allow to draw any conclusion about
non-perturbative physics. Calculations of missing higher order terms
including the mass effects certainly would help a lot to draw more
reliable conclusions about the borderline of validity of pQCD.

\bigskip
\newpage

{\bf Acknowledgments } 
One of us (A.L.K) is grateful to D. V. Shirkov for his interest in the
present work and for useful discussions. He (A.L.K) wishes to thank
DESY Zeuthen where part of this work was done for its hospitality.\\


\appendix{\large \bf Appendix}

Here we present the explicit formulae for the two--loop vacuum polarization 
amplitudes of the flavor diagonal vector current.
We first introduce the abbreviations
\bea
f = -\frac{1}{2} \ln \xi\,,\;\;
g =\ln (1-\xi)\,,\;\;
h = \ln (1+\xi) \nn
\eea
where $y$ and $\xi$ are given by (\ref{xivar})
and
\bea
\Delta {\rm Li}_3 = 2{\rm Li}_3(\xi)-{\rm Li}_3(\xi^2)\;,\;\;\;
\Delta {\rm Li}_2 = {\rm Li}_2(\xi)-{\rm Li}_2(\xi^2) \nn
\eea
and define the auxiliary functions
\bea
{\rm XX} &=& \Delta {\rm Li}_3+\frac{8}{3}\;f\;\Delta {\rm Li}_2+\frac{4}{3}\;
 f^2\;(2h+g)\;, \crn
{\rm YY} &=& \frac{8}{3}\;(\Delta {\rm Li}_2+2f\;(2h+g)+3f^2)\;\;.    \nn
\eea
The poly-logarithms
${\rm Li}_n(z)$ are defined by the integrals
\begin{eqnarray*}
{\rm Li}_n(z)=\frac{(-1)^{n-1}}{(n-2)!}\int_0^1 
\frac{\ln^{n-2}(x)\:\ln(1-xz)}{x}\:dx 
\end{eqnarray*}
and $\zeta_n=\zeta(n)=\sum_{i=1}^{\infty}1/n^i$ is the Riemann 
zeta--function, with values $\zeta_2=\pi^2/6$, $\zeta_3=1.202057...$, 
$\zeta_4=\pi^4/90$ etc.. 

In terms of these functions we find the following compact expression
for the two-loop contribution to the Adler function related to the QCD
vector neutral current vacuum polarization amplitude:\footnote{ it is
given by
$$\dot{\Pi}^{'(2)}_V=\frac{1}{3}\,\left({\mathrm{Re}}\Pi^{NC}_V(s)/s
-{\mathrm{Re}}\dot{\Pi}^{NC}_V(s)\right)$$ in terms of the amplitudes
which we can find in Sec.~4.5 of Ref.~\cite{FR} (see
also~\cite{ChGvN}). There the rules for the analytic continuation
to the time--like region can also be found.}
\bea 
12\pi^2\,\dot{\Pi}^{'(2)}_V&=&
-2y^2\,{ \rm XX} 
+\sqrt{1-y}\,\left( (1+y)\,{ \rm YY} - 14/3\, y\,f\, \right) \crn
 & &              +\left(3y^2-4-4/(1-y)\right)\,f^2   \crn
 & &              +1+11/3\,y+2y^2\,\zeta(3)\;.  
\eea

The following exercise about handling asymptotic expansions is
intended to illustrate at the two--loop level the procedure to be
applied at the three--loop level where an exact massive result is
not yet known.\\

The expansion for large $Q^2$ $(y=-4m^2/Q^2 \ra 0)$ reads ($\:{\rm L}\equiv
\ln (Q^2/m^2)\:$)
\bea 
12\pi^2\,\dot{\Pi}^{'(2)}_V&=&
~~1
+3\,(1-{\rm L})\,y
+\,({{17}\over{24}}
+2\,\zeta(3)
+{{1}\over{4}}\,{\rm L}
-{{3}\over{2}}\,{\rm L}^{2})\,y^{2} \crn
&+&({{139}\over{108}}
+{{19}\over{9}}\,{\rm L}
-{{29}\over{24}}\,{\rm L}^{2})\,y^{3}
+({{2173}\over{13824}}
+{{2575}\over{1152}}\,{\rm L}
-{{203}\over{192}}\,{\rm L}^{2})\,y^{4} \crn
&+&(-{{8311}\over{25600}}
+{{75727}\over{34560}}\,{\rm L}
-{{1169}\over{1152}}\,{\rm L}^{2})\,y^{5}
+(-{{784577}\over{1382400}}
+{{100063}\over{46080}}\,{\rm L}
-{{383}\over{384}}\,{\rm L}^{2})\,y^{6}\crn
&+& \cdots \;\;,
\eea
and for low $Q^2$ we have
\bea 
12\pi^2\,\dot{\Pi}^{'(2)}_V&=&
-({{328}\over{81}}\,y^{-1}
+{{3592}\over{675}}\,y^{-2}
+{{999664}\over{165375}}\,y^{-3} 
+{{831776}\over{127575}}\,y^{-4}
+{{1729540864}\over{252130725}}\,y^{-5} \crn
&+&{{43321977728}\over{6087156075}}\,y^{-6} 
+{{401009026048}\over{54784404675}}\,y^{-7}
+{{5064168384512}\over{676610809875}}\,y^{-8}
+ \cdots \:)
\eea

Both series expansions diverge at the boundary of the circle of
convergence $Q^2=4m^2$. In fact this is precisely in the region where
mass effects are of the order of unity in the Euclidean region. We
therefore apply a conformal mapping
\bea
y^{-1}=\frac{-Q^2}{4m^2} \ra \omega = \frac{1-\sqrt{1-1/y}}{1+\sqrt{1-1/y}}
\eea

\rput{90}(10.0,-3.4){\scalebox{0.7 0.7}{%
\epsfbox{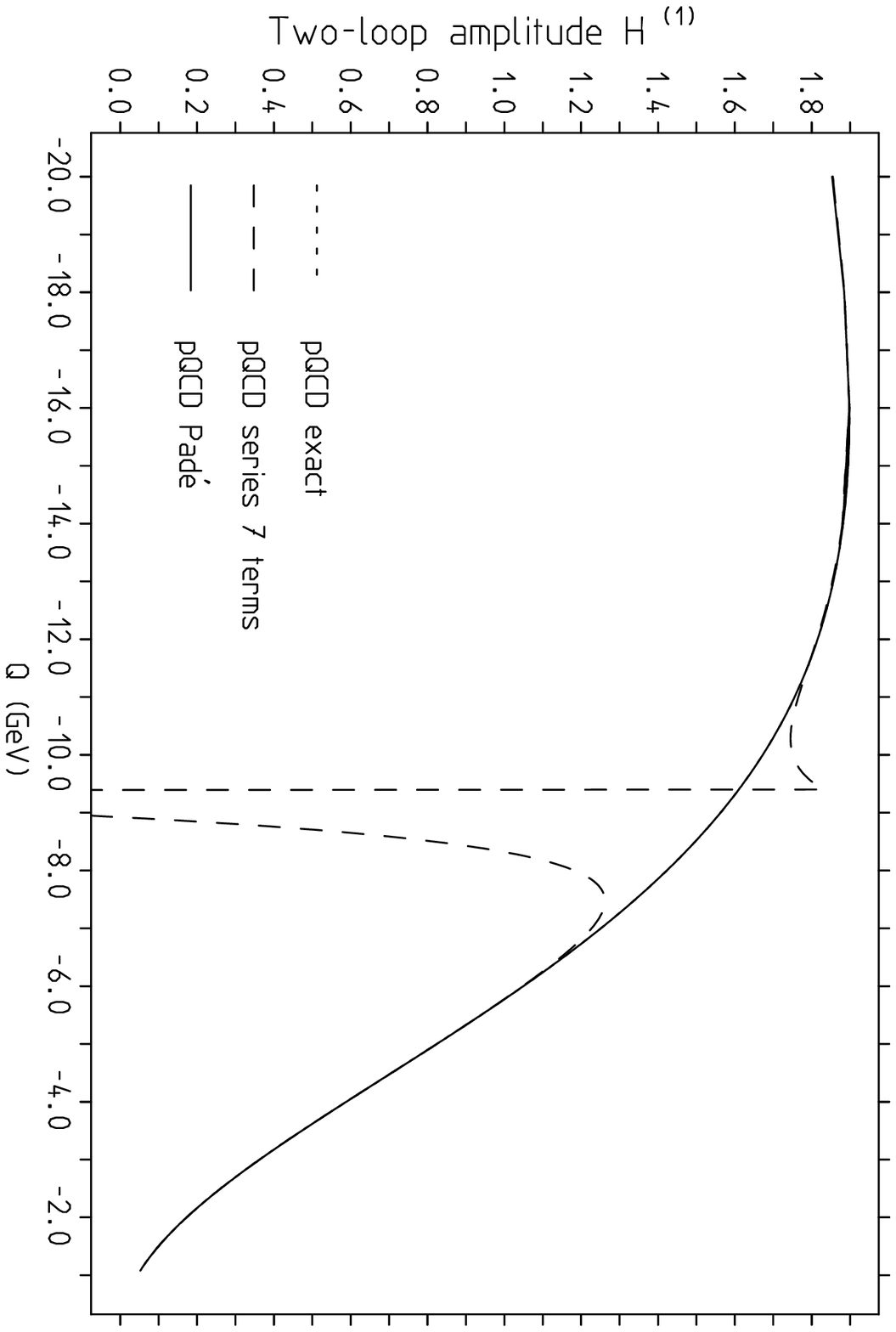}}}

\vspace*{8.5cm}

\begin{center}
\begin{minipage}[h]{15.2cm} \baselineskip 12truept \noindent
\small{{Figure~3:
The two--loop amplitude $H{(1)}=(12\pi^2)\dot{\Pi}^{'(2)}_V$: shown is
the exact result (dotted line), the series expansions for low $Q^2$ (8
terms) ($4m^2>Q^2$) and for large $Q^2$ (7 terms) ($Q^2>4m^2$) (dashed
line) and the Pad\'e improvement (full line) of the low $Q^2$ series
working excellent up to ($16m^2>Q^2$). For higher $Q^2$ the large
$Q^2$--expansion works perfectly. The dotted line is not seen because
of the perfect agreement between the exact result and the Pad\'e
improved approximation.}}
\end{minipage}
\end{center}

from the complex negative $q^2$ half--plane to the interior of the
unit circle $|\omega|<1$ together with Pad\'e resummation as proposed
in~\cite{FT}. The Pad\'e approximant provides a good estimation to
much higher values of $1/y$ up to about $1/y\sim 4$. This is displayed in
Fig.~3, which shows that utilizing the Pad\'e improvement allows us
to obtain reliable results also in the relevant Euclidean
``threshold region'', around $y=1$. \\

The three--loop vector current amplitude $\Pi^{'(3)}$ is available in
the form of a small $Q^2$ expansion, Eq. (47) of Ref.~\cite{ChKSsQ}, and as
a large $Q^2$ expansion, Eqs.~(7-10) of Ref.~\cite{ChHKShQ}. By the
definition (\ref{Adef}) the contribution to the Adler function is
obtained by differentiation
\bea
\dot{\Pi}^{'(3)}(y)=-y \, d\Pi^{'(3)}(y)/dy\;\;.
\eea
While the low $Q^2$ expansion is given in the on-shell scheme, where
it is a simple power series in $1/y$, the high $Q^2$ expansion is given
in the \MSb scheme as an expansion in $\bar{y}=4\bar{m}^2/Q^2$ where
$\bar{m}=\bar{m}(\mu^2)$ is the \MSb running mass at scale $\mu$ and
the \MSb amplitude $\bar{\Pi}^{'(3)}$ 
includes powers of the logarithms $l_{qm}\equiv\ln(-q^2/\bar{m}^2)$
and $l_{q\mu}\equiv\ln(-q^2/\mu^2)$ in additions to powers of
$\bar{y}$. \\

Since, for simplicity, we work with
pole masses, we have to reparametrize the \MSb amplitude.
The relationship between the \MSb mass
$\bar{m}$ and the pole mass $m$ is given by~\cite{ChKSsQ,GBGS,FJTV}
\begin{equation}
m=\bar{m} \left( 1+c_1 \left( \frac{\alpha_s}{4 \pi} \right)
     +c_2 \left( \frac{\alpha_s}{4 \pi} \right)^2 + \ldots \right),
\lab{Mm}
\end{equation}
with
\begin{eqnarray}
&&c_1=C_F(4+3L), \\
&& \nonumber \\
&&c_2=
 C_FC_A \left(\frac{1111}{24}-8\zeta_2-4I_3(1)+\frac{185}{6}L
 +\frac{11}{2}L^2\right)
  \nn \\
&& \nn \\
&&~~~
 -C_F T_F n_f \left(\frac{71}{6}+8\zeta_2+\frac{26}{3}L+2L^2 \right)
  \nn \\
&& \nn \\
&&~~~
+C_F^2\left(\frac{121}{8} +30\zeta_2+8I_3(1)
 +\frac{27}{2} L+\frac92 L^2 \right)-12C_F T_F \left(1-2\zeta_2 \right).
\end{eqnarray}
where 
$I_3(1)=\frac{3}{2} {\zeta}_3 - 6{\zeta}_2 \ln 2$, $L=\ln(\mu^2/m^2)$,
$C_A=3$, $C_F=4/3$, $T_F=1/2$ and $n_f=6$.

Asymptotically, for large $Q^2$ we find ($l_{q\mu}\equiv\ln(-q^2/\mu^2)$)
\bea
H^{(2)}(\infty)&=&\frac{1}{32}\left\{
(-3\,C_F^2+(123-88\,\zeta_3-22\,l_{q\mu})\,C_A C_F \right. \crn
&+& \left. (-44+32\,\zeta_3+8\,l_{q\mu})\,C_F T_F n_l
+(-44+32\,\zeta_3+8\,l_{q\mu})\,C_F T_F)\right\}
\eea
which coincides with the \MSb expression 
for the $\alpha_s^2$ coefficient in the $D$-function, calculated in
Ref.\cite{ChKT}.
At the scale $\mu^2=Q^2$ is of order unity:
$H^{(2)}(\infty)\simeq 1.40923$. 

Due to the logs $\ln(Q^2/m^2)$, still present after setting
$\mu^2=Q^2$ in the three--loop amplitudes, we perform the Pad\'e
improvement for the non-log and the log terms separately. The result
is depicted in Fig.~4.

Note that the Pad\'e approximant
is an analytic continuation of the small $Q^2$ expansion with 8 terms.
From a practical point of view it provides the ``exact'' mass
dependence of the three--loop amplitude.

\rput{90}(10.0,-3.0){\scalebox{0.7 0.7}{%
\epsfbox{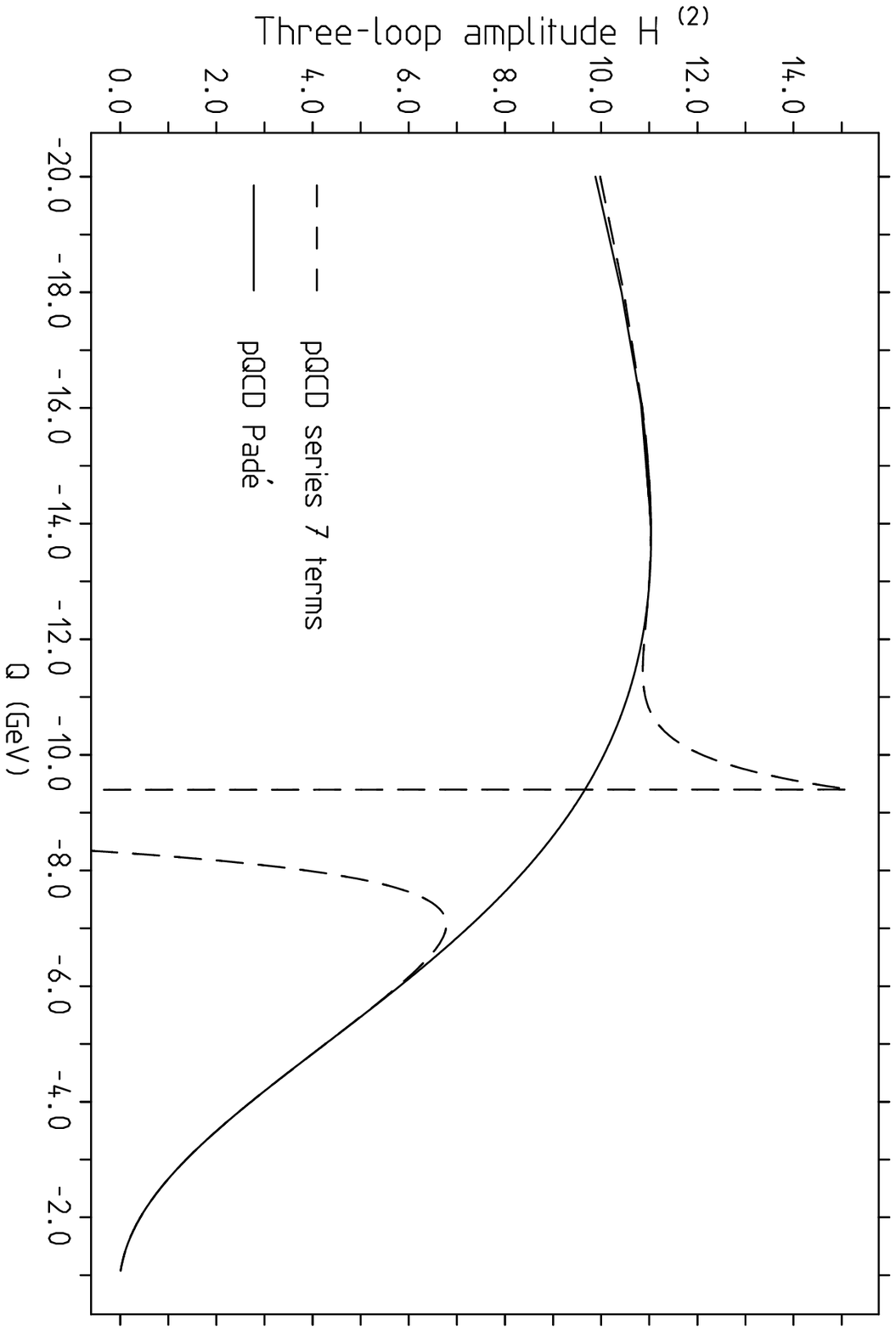}}}

\vspace*{8.5cm}

\begin{center}
\begin{minipage}[h]{15.2cm} \baselineskip 12truept \noindent
\small{{Figure~4:
The three--loop amplitude $H^{(2)}=(12\pi^2)\dot{\Pi}^{'(3)}_V$: shown
are the series expansions for low $Q^2$ (8 terms) ($4m^2>Q^2$) and for
large $Q^2$ (7 terms) ($Q^2>4m^2$) (dashed line) and the Pad\'e
improvement (full line) of the low $Q^2$ series working excellent up
to ($16m^2>Q^2$). For higher $Q^2$ the large $Q^2$--expansion works
perfectly.}}
\end{minipage}
\end{center}


\bb{99}
                                              
\bibitem{EJ}
S.I.~Eidelman and F.~Jegerlehner, {\it Z. Phys.} {\bf C 67} (1995) 585;\\
F.~Jegerlehner, {\it Nucl. Phys. B} (Proc. Suppl.) {\bf 51C} (1996) 131.

\bibitem{JT}
F.~Jegerlehner and O.V.~Tarasov, DESY~98-093 (hep-ph/9809485) and
references therein.

\bibitem{ChKSsQ} 
K.G. Chetyrkin, J.H. K\"uhn and  M. Steinhauser, {\it Nucl.
Phys.} {\bf B 482} (1996) 213.  

\bibitem{ChHKShQ} 
K.G. Chetyrkin, R. Harlander, J.H. K\"uhn and  M. Steinhauser, 
{\it Nucl. Phys. } {\bf B 503} (1997) 339.

\bibitem{PQW} 
H. Poggio, H. Quinn and S. Weinberg,  {\it Phys. Rev.} {\bf D 13}
(1976) 1958.

\bibitem{Barnet}
R. Barnett, M. Dine  and L. McLerran, {\it Phys. Rev.} {\bf D 22}
(1980) 594.

\bibitem{SVZ}
M.A. Shifman, A.I. Vainshtein and V.I. Zakharov, {\it Nucl.Phys.} 
{\bf B 147} (1979) 385. 

\bibitem{DeRafael}
R. Bertelman, G. Launer and E. De Rafael, {\it Nucl. Phys.} 
{\bf B 250} (1985) 61. 

\bibitem{CKT}
K.G. Chetyrkin, N.V. Krasnikov  and A.N. Tavkhelidze,  
{\it Phys. Lett.} {\bf 76B} (1978) 83.

\bibitem{Shankar}
R. Shankar, {\it Phys. Rev.} {\bf D 15} (1977) 755.

\bibitem{E}
S.I. Eidelman, L.M. Kurdadze and A.I. Vainshtein,
{\it Phys.Lett.} {\bf 82B} (1979) 278.

\bibitem{PDG} 
C. Caso et al.(Particle Data Group), {\it Eur. Phys. J.} {\bf C 3} (1998) 1.

\bibitem{ChKT}
K.G. Chetyrkin, A.L. Kataev and F.V. Tkachov, {\it Phys. Lett.}
{\bf 85B} (1979) 277;\\
M. Dine  and J. Sapirstein, {\it Phys. Rev. Lett.} {\bf 43} (1979) 
668;\\
W. Celmaster and R.J. Gonsalves, {\it Phys. Rev. Lett.} {\bf 44} (1980)
560.

\bibitem{GKL}
S.G. Gorishny, A.L. Kataev and S.A. Larin, {\it Phys. Lett.} {\bf B 259} (1991)
144;\\
L.R. Surguladze and M.A. Samuel, {\it Phys. Rev. Lett.} {\bf 66} 
   (1991) 560; ibid. 2416 (Err),\\
K.G. Chetyrkin, {\it Phys. Lett.} {\bf B 391} (1997) 402.

\bibitem{GKLM}
S.G. Gorishny, A.L. Kataev and S.A. Larin, {\it Nuov. Cim.} {\bf 92A}
(1986) 119.

\bibitem{Adler}
S. Adler, {\it Phys. Rev.} {\bf D 10} (1973) 3714.

\bibitem{RG}
A. De Rujula  and H. Georgi, {\it Phys. Rev.} {\bf D 13} (1976) 1296.

\bibitem{AK}
A.L. Kataev,  Invited talk at the II Workshop "Continuous 
Advances in QCD", Minneapolis, March 1996, 
World Scientific 1996, ed. M.I.Polikarpov, p.107.
(hep-ph/9607426).

\bibitem{Raf}
S. Peris, M. Perrottet and E. de Rafael,  {\it J. High Energy Phys.} 
{\bf 05} (1998) 011.

\bibitem{ChK}
K.G.~Chetyrkin and J.H.~K\"uhn, {\it Phys. Lett.} {\bf B 342} (1995) 356.

\bibitem{FT}
J.~Fleischer and  O.V.~Tarasov, {\it Z. Phys.} {\bf C 64} (1994) 413.

\bibitem{NPho}
D.J. Broadhurst and S.C. Generalis, Open University preprint OUT-4102-12 (1984)
(unpublished); S.C. Generalis, {\it J. Phys.} {\bf G 15} (1989) L225.

\bibitem{ChGS}
K.G. Chetyrkin, S.G. Gorishny and V.P. Spiridonov, {\it Phys. Lett.} {\bf
160B} (1985) 149.

\bibitem{SCh}
V.P. Spiridonov and K.G. Chetyrkin, {\it Yad. Fiz.} {\bf 47} (1988) 818 
[{\it Sov. J. Nucl. Phys.} {\bf 47} (1988) 522]

\bibitem{ST90}
L.R.~Surguladze and F.V.~Tkachov, {\it Nucl. Phys.} {\bf B 331}
(1990) 35.

\bibitem{Nar89}
S. Narison, ''QCD Spectral Sum Rules'', (Lecture Notes in Physics,
Vol. 26), World Scientific, Singapore 1989.

\bibitem{DoNa}
H.G. Dosch and S. Narison, {\it Phys. Lett.} {\bf B 417} (1998) 173.

\bibitem{Gasser}
J. Gasser, and H. Leutwyler, {\it Phys. Rep.} {\bf C 87} (1982) 77;\\
J. Bijnens, J. Padres and E. de Rafael, {\it Phys. Lett.} {\bf B 348}
(1995) 226;\\
H. Leutwyler, {\it Phys. Lett.} {\bf B 378} (1996) 313.

\bibitem{LEP} 
D. Karlen, {\em Experimental Status of the Standard Model},
talk presented at the \\  ``XXIX International Conference on
 High-Energy Physics'' (ICHEP 98), \\ Vancouver 1998. 

\bibitem{FR}
F. Jegerlehner, {\it Prog. Part. Nucl. Phys.} {\bf 71} (1991) 1, Sec.4.5.

\bibitem{Decoupling} 
W.~Bernreuther and  W.~Wetzel, {\it Nucl. Phys.} {\bf B 197} (1983) 228; 
ibid.
{\bf B 513} (1998) 758 (Err);\\
W.J.~Marciano, {\it Phys. Rev.} {\bf D 29} (1984) 580;\\
S.A. Larin, T. van Ritbergen and J.A.M. Vermaseren, 
{\it Nucl. Phys.} {\bf B 438} (1995) 278;\\
K.G. Chetyrkin, B.A. Kniehl and M. Steinhauser, {\it Phys. Rev. Lett.} 
{\bf 74} (1997) 2184; \\
G. Rodrigo, A. Pich, and A. Santamaria,
{\it Phys. Lett.} {\bf B 424} (1998) 367.

\bibitem{BvN} 
J. Bl\"umlein and W.L. van Neerven, DESY-98-176, hep-ph/9811351.

\bibitem{ChGvN}
T.H.Chang, K.J.F.Gaemers and W.L. van Neerven, {\it Nucl. Phys. }
{\bf B 202} (1982) 407; \\
L.J. Reinders, H.R. Rubinstein and S. Yazaki, {\it Phys. Rep.} {\bf C 127}
(1985) 1; \\
B.A. Kniehl, {\it Nucl. Phys. } {\bf B 347} (1990) 86.

\bibitem{GBGS} 
N.~Gray, D.J. Broadhurst, W. Grafe and K. Schilcher, {\it Z. Phys.} 
{\bf C 48} (1990) 673.

\bibitem{FJTV}
J.~Fleischer, F.~Jegerlehner, O.V.~Tarasov and  O.L.~Veretin, 
DESY~98-026, hep-ph/9803493, {\it Nucl. Phys.} {\bf B 539} (1999) to appear.

\end{thebibliography}

\end{document}